\documentclass[review]{elsarticle}
\usepackage[top=3cm,bottom=3cm,left=3cm,right=3cm]{geometry}
\usepackage{lipsum}

\usepackage{lineno,hyperref}
\usepackage{textcomp}
\usepackage{amsfonts}
\usepackage{amsmath}
\usepackage{amssymb}
\usepackage{rotfloat}
\biboptions{numbers,sort&compress}
\usepackage{graphicx}% Include figure files
\usepackage{dcolumn}% Align table columns on decimal point
\usepackage{bm}% bold math
\usepackage{array}
\usepackage{setspace} 
\usepackage{bm}% bold math
\usepackage{array}
\usepackage{setspace}
\usepackage{xcolor}

\journal{Jounal of Power Sources}

\begin{document}

\begin{frontmatter}

\title{Response to ``Comment on `Origin of the Curie--von Schweidler law and
the fractional capacitor from time-varying capacitance {[}J.~Power~Sources~532~(2022)~231309{]}' ''}

\author{Vikash Pandey}
\address{School of Interwoven Arts and Sciences, Krea University, Sri City, India}

\ead{vikash.pandey@krea.edu.in}

\begin{abstract}
We welcome Allagui~et~al.\textquoteright s discussions about our recent paper that has proposed revisions to the existing theory of
capacitors. It gives us an opportunity to emphasize on the physical
underpinnings of the mathematical relations that are relevant for
modeling using fractional derivatives. The concerns
raised by Allagui~et~al. are found to be quite questionable when examined in light
of the \textit{established} standard results of fractional calculus. Consequently,
the inferences that they have drawn are not true. Finally,
we would like to thank Allagui~et~al. for their Comment because this subsequent Response has actually led to a further consolidation of our
results that are supposed to be significant for materials science
as well as for fractional control systems and engineering.  
\end{abstract}


\end{frontmatter}

%\linenumbers

\textbf{NOTE:} \texttt{The peer-reviewed version of this Response is published in
the Journal of Power Sources, Vol.~551, 232167, 2022.}\\
DOI: 
\href{https://doi.org/10.1016/j.jpowsour.2022.232167}{https://doi.org/10.1016/j.jpowsour.2022.232167}. 

\texttt{The published version of the manuscript is available online
at 
\\\href{https://www.sciencedirect.com/science/article/pii/S0378775322011442}{https://www.sciencedirect.com/science/article/pii/S0378775322011442}.}

\textcolor{blue}{\textbf{This document is an e-print which may differ
in, e.g. pagination, referencing styles, figure sizes, and typographic
details.}}

\newpage

In our recent publication \cite{Pandey2022a},  we gave a physical interpretation
to the century-old Curie--von Schweidler law of dielectrics which
interestingly may also be seen as an electrical analogue of Nutting's law from rheology \cite{Pandey2016a,Pandey2016b}.
The derivation of the Curie--von Schweidler law required a revision
of the classical charge--voltage relation of capacitors, $Q\left(t\right)=C_{0}V\left(t\right)$,
where $Q$ is the charge, $C_{0}$ is the constant
capacitance of the capacitor, $V$ is the voltage, and $t$ is the time. It is worthwhile to note that the classical
relation is more than two-centuries old \cite{Keithley1999} and it was originally proposed for a constant capacitance capacitor. In contrast most capacitors of practical applications exhibit a time-varying capacitance but in lack of a better relation the same-old classical relation has been inadvertently used. In order to overcome this lacking we proposed a new charge--voltage relation which is equally applicable for a constant capacitance capacitor as well as for a capacitor with a time-varying capacitance. The relation is expressed by Eq.~$\left(9\right)$ in our paper as \cite{Pandey2022a}:
\begin{equation}
Q\left(t\right)=C\left(t\right)\ast\dot{V}\left(t\right),\label{eq:convolv}
\end{equation}
where $\ast$ represents the convolution operation and $C\left(t\right)$ is the time-varying capacitance of the capacitor. In their Comment on our paper, Allagui~et~al. \cite{Allagui2022} have raised certain concerns with the use of the convolution integral of linear time-varying capacitance with the time-derivative of voltage. In this Response, for the sake of clarity, completeness, and correctness, we prove that each of their observations and the inferences that they have drawn from them are quite questionable and therefore not valid.

\begin{enumerate}

\item According to Allagui~et~al. capacitances being added in parallel
is a frequency domain assertion, and that they cannot be added in the time domain \cite{Allagui2022}. Moreover,
they have the opinion that circuit synthesis is not done in the time domain.

\textit{Response:} We would like to emphasize that as long as the linearity of the system holds, circuit synthesis could be done in the time domain as well as in the frequency domain. We transform a problem from one domain to another domain depending upon the domain in which finding a solution to the problem is relatively easier. For example, modeling a digital filter is often easier in the frequency domain as it facilitates a convenient framework for the  mathematical analysis of its properties. In contrast, a sampling process is meaningful in the
time domain. The two domains should be seen as the two different viewpoints from which to understand the problem. It is also worthwhile to respect the fact that the time domain is the domain of physical reality and it is probably more
intuitive than the frequency domain.

It has already been shown that our expression for current matches exactly with the predictions from Kirchhoff's current
law for capacitors \cite{Pandey2022a}. Since Kirchhoff's law stems from the fundamental principle of conservation
of charge, this further gives confidence in all the mathematical relations that have been proposed
and validated in our work. We again prove it here through a simple example in which a linearly time-varying
voltage is applied to a capacitor with a linearly time-varying capacitance. If we assume, $V\left(t\right)=at$, $C\left(t\right)=C_{0}+\phi t$, such that,  $a>0$ and $\phi>0$, then the classical relation predicts:
\begin{equation}
Q\left(t\right)=C\left(t\right)V\left(t\right)=\left(C_{0}+\phi t\right)\cdot at=C_{0}at+\phi at^{2},\label{eq:classical_charge}
\end{equation}
from which the respective current, $I$, is obtained after differentiating Eq.~$\left(\ref{eq:classical_charge}\right)$
as:
\begin{equation}
I\left(t>0\right)=C_{0}\cdot a+2\phi t\cdot a. \label{eq:classical_current}
\end{equation}
Since, $a=\dot{V}\left(t\right)$, the two terms on the right
hand side of Eq.~$\left(\ref{eq:classical_current}\right)$ may be interpreted as two capacitive
currents that flow through two capacitors arranged in a parallel combination,
the first of which has a constant capacitance, $C_{0}$, and the
second capacitor has a time-varying capacitance, $2\phi t$. This means that according to Eq.~$\left(\ref{eq:classical_current}\right)$, the capacitance of the capacitor is, $C\left(t\right)=C_{0}+2\phi t$, which contradicts our initial assumption of a capacitance, $C\left(t\right)=C_{0}+\phi t$. To summarize,
the classical relation yields an \textit{additional capacitor} of
capacitance, $\phi t$, which was not initially present in our chosen example. Since this inconsistency that is observed from Eq.~$\left(\ref{eq:classical_current}\right)$ has its
origin in Eq.~$\left(\ref{eq:classical_charge}\right)$, it is evident that the classical relation is not valid for capacitors with a time-varying capacitance. This is because the classical relation dictates a term-by-term multiplication
of $C\left(t\right)$ and $V\left(t\right)$ which is applicable for a time-invariant system. On the contrary, a capacitor with a time-varying capacitance constitutes a time-variant system.

Now, we extract the correct results from the proposed convolution relation,
Eq.~$\left(\ref{eq:convolv}\right)$, that is also in accordance with the established
laws of current-electricity. Evaluating the convolution of $C\left(t\right)$ and $\dot{V}\left(t\right)$ in time-domain, we have: 
\begin{equation}
Q\left(t\right)=C\left(t\right)\ast\dot{V}\left(t\right)=a\int\limits _{0}^{t}\left[C_{0}+\phi t\right]d\tau=C_{0}at+\frac{\phi}{2}at^{2}.\label{eq:new_charge}
\end{equation}
On comparing Eq.~$\left(\ref{eq:classical_charge}\right)$ with Eq.~$\left(\ref{eq:new_charge}\right)$, we observe that both the classical relation and the convolution relation yield a similar quadratic dependence of charge on time but they are not exactly identical. As expected, this difference also reflects from the respective expression for the current which is,
\begin{equation}
I\left(t>0\right)=C_{0}\cdot a+\phi t\cdot a.\label{eq:new_current}
\end{equation}
It is seen that Eqs.~$\left(\ref{eq:new_charge}\right)$ and $\left(\ref{eq:new_current}\right)$ obtained from
the convolution relation are different from their respective counterparts,
Eqs.~$\left(\ref{eq:classical_charge}\right)$ and $\left(\ref{eq:classical_current}\right)$, that are obtained from the classical relation. The result
obtained from the convolution relation has two distinct merits over
the result obtained from the classical relation. First, since, $a=\dot{V}\left(t\right)$,
the two terms on the right hand side of Eq.~$\left(\ref{eq:new_current}\right)$ may
be interpreted as two capacitive currents that flow through two capacitors
arranged in a parallel combination, the first of which has a constant-capacitance, $C_{0}$, and the second capacitor has a time-varying
capacitance, $\phi t$. It is emphasized that we did not encounter the \textit{additional capacitor} here which had mysteriously emerged from  Eq.~$\left(\ref{eq:classical_current}\right)$. Second, the application of Kirchhoff's current law for capacitors yields exactly the
same equation as expressed by  Eq.~$\left(\ref{eq:new_current}\right)$. This agreement with Kirchhoff's current law provides a physical validity to the convolution relation, Eq.~$\left(\ref{eq:convolv}\right)$. 
\\
\item Next, Allagui~et~al. claim that our proposed convolution relation, i.e.,
Eq.~$\left(\ref{eq:convolv}\right)$, generates a constant current and hence a zero charge if the input is a step voltage \cite{Allagui2022}.

\textit{Response:} We refer the readers to the definition of a fractional derivative which is expressed by Eq.~$\left(2\right)$ in
Ref.~\cite{Pandey2022a} and the sentence that precedes the respective equation. We reiterate
that particular excerpt from the paper here.

``The Caputo definition
for the fractional derivative of a \textit{causal}, continuous function,
$f\left(t\right)$, is the convolution of an integer-order derivative
with a power-law memory kernel, $\phi_{\alpha}\left(t\right)$, as
Ref.~{[}33{]}:
\begin{equation}
\frac{d^{\alpha}}{dt^{\alpha}}f\left(t\right)\overset{\text{def.}}{=}\dot{f}\left(t\right)\ast\phi_{\alpha}\left(t\right),\text{ }\phi_{\alpha}\left(t\right)=\frac{t^{-\alpha}}{\Gamma\left(1-\alpha\right)},\text{ }0<\alpha<1,\label{eq:frac_deriv}
\end{equation}
where the number of over-dots represent the order of differentiation
with respect to time, $t$.''  Rewriting this expression in its
respective integral form with its generic limits of integration, we
have:
\begin{equation}
\frac{d^{\alpha}}{dt^{\alpha}}f\left(t\right)\overset{\text{def.}}{=}\frac{1}{\Gamma\left(1-\alpha\right)}\int\limits _{-\infty}^{t}\frac{\dot{f}\left(\tau\right)}{\left(t-\tau\right)^{\alpha}}d\tau ,\text{ }0<\alpha<1.\label{eq:convolv_integ}
\end{equation} 
Further, we draw attention to the standard results mentioned in Chapter~1, in particular on p.~16 of the textbook \cite{Mainardi2010} which emphasizes on \textit{causal} functions and the care required in solving convolution-integrals that involve them. If $f\left(t\right)$ is a causal function, it necessitates
$f\left(t\right)=0$ at all times, $t<0$. Further, if there is a finite jump discontinuity of the integrand at $t=0$, the following holds: 
\begin{equation}
\int\limits _{-\infty}^{t}\left(\cdots\right)d\tau=\int\limits _{0^{-}}^{t}\left(\cdots\right)d\tau.\label{eq:integrand}
\end{equation}
In the case of, $0<\alpha<1$, following Eq.~$\left(\ref{eq:integrand}\right)$, the right hand side of Eq.~$\left(\ref{eq:convolv_integ}\right)$ is decomposed in accordance with the following identity:
\begin{equation}
\frac{1}{\Gamma\left(1-\alpha\right)}\int\limits _{0^{-}}^{t}\frac{\dot{f}\left(\tau\right)}{\left(t-\tau\right)^{\alpha}}d\tau=\frac{f\left(0^{+}\right)}{\Gamma\left(1-\alpha\right)}t^{-\alpha}+\frac{1}{\Gamma\left(1-\alpha\right)}\int\limits _{0}^{t}\frac{\dot{f}\left(\tau\right)}{\left(t-\tau\right)^{\alpha}}d\tau.\label{eq:saviour}
\end{equation}
Since Allagui~et~al.\textquoteright s concerns are centered on the term, $\dot{V}\left(t\right)$, independent of the nature of the capacitance, $C\left(t\right)$, in Eq.~$\left(\ref{eq:convolv}\right)$,
 for simplicity
we assume the capacitor has a constant capacitance, $C_{0}$. Mathematically, this means the memory kernel, $\phi_{\alpha}\left(t\right)$, is scaled by $C_{0}$, and $\alpha\rightarrow 0$. Identifying, $f$ as $V$, we express the convolution relation, Eq.~$\left(\ref{eq:convolv}\right)$,
in light of Eqs.~$\left(\ref{eq:frac_deriv}\right)$--$\left(\ref{eq:saviour}\right)$, as:
\begin{equation}
Q\left(t\right)=C_{0}V\left(0^{+}\right)+C_{0}\int\limits _{0}^{t}\dot{V}\left(\tau\right)d\tau\label{eq:saviour_prov}.
\end{equation}
On imposing Allagui~et~al.\textquoteright s condition, $\dot{V}\left(t\right)=0$, i.e., $V\left(t\right)=V_{0}\geq0$, for $t\geq0$, where $V_{0}$
is the constant voltage applied to the capacitor, the integral-term in Eq.~$\left(\ref{eq:saviour_prov}\right)$ vanishes, leaving behind,
\[
Q\left(t\right)=C_{0}V\left(0^{+}\right)=C_{0}V_{0}\neq 0, \text{ if, } C_{0}\neq 0 \text{, and } V_{0}\neq 0.
\]
If the voltage
at time, $t=0$, is, $V_{0}=0$, then the charge at all times is zero as long as, $\dot{V}\left(t\right)=0$, is maintained. This is expected from a discharged capacitor. In contrast, if the voltage at time, $t=0$, is, $V_{0}>0$,
then the charge at all times is $C_{0}V_{0}$ as long as, $\dot{V}\left(t\right)=0$,
is maintained. This is expected from a charged capacitor. These results match exactly with those obtained from the classical charge--voltage relation for cases in which a constant voltage is fed to a constant-capacitance capacitor. 

In the case of a capacitor characterized with a time-varying capacitance, i.e., $0<\alpha<1$,  if the voltage, $V\left(t\right)=V_{0}>0$, and $\dot{V}\left(t\right)=0$, for $t\geq0$, the corresponding last term from Eq.~$\left(\ref{eq:saviour}\right)$ vanishes. Consequently, the resulting charge is obtained from the corresponding non-integral term of Eq.~$\left(\ref{eq:saviour}\right)$, which in this case turns out to be, $Q\left(t\right) \propto t^{-\alpha}$. Clearly, neither the charge is zero nor the current is a constant in such a case. This is in accordance with the experimental observations but contrary to the incorrect inferences drawn by Allagui.~et~al. Therefore, the proposed convolution relation, Eq.~$\left(\ref{eq:convolv}\right)$, completes the bigger picture and yet retains the significance of the classical relation for constant-capacitance capacitors.
\\
\item Allagui~et~al. claim \cite{Allagui2022} that the question that we raised about the dimensional
consistency of their equations in Ref.~\cite{Fouda2020} is not correct.

\textit{Response:} Although a detail discussion on this can be found in Ref.~\cite{Pandey2022b}, we briefly mention it here for the completeness of this Response. To summarize, the authors of Ref.~\cite{Fouda2020} seem to have given conflicting information about the same symbol used in their paper. We reproduce the relevant excerpts from their paper here. First, they mention in the paragraph after Eq.~$\left(4\right)$:

\textit{``By applying the principle of causality, the operation by which
the charge $q\left(t\right)$ is created in the time-domain is therefore a convolution
operation of capacitance $c\left(t\right)$ and voltage $v\left(t\right)$
such that $q\left(t\right)=c\left(t\right)\ast v\left(t\right)$ which
is contrary to the usual assumption of a multiplication operation [19].''}

It is evident that the authors have \textit{declared} $c\left(t\right)$ as the capacitance, and $v\left(t\right)$ as the voltage, but the unit of charge, $q\left(t\right)$, according to their relation, $q\left(t\right)=c\left(t\right)\ast v\left(t\right)$, turns out to be of the quantity, \textit{$\mbox{capacitance}\cdot \mbox{voltage} \cdot \mbox{time}$}, which is, $\mbox{Coulomb}\cdot \mbox{s}$. Since the correct unit for charge is $\mbox{Coulomb}$, their proposed equation is dimensionally inconsistent which we had indicated in Ref.~\cite{Pandey2022a}. Later, the authors have mentioned after Eq.~$\left(5\right)$:

\textit{``For ideal
capacitors, the capacitance is considered to be a geometric constant
independent of the applied frequency which implies that in the time-domain
$c\left(t\right)=C\delta\left(t\right)$, which leads to $q\left(t\right)=C\delta\left(t\right)\ast v\left(t\right)=Cv\left(t\right)$.''}

We know that the unit of the Dirac-delta function is inverse of its argument which implies that in the time domain it is $\mbox{1/s}$. Since Allagui~et~al. had already declared, $c$, as the capacitance, it is therefore inferred that, $C=c\left(t\right)/\delta\left(t\right)$, has the unit of  $\mbox{Farad}\cdot\mbox{s}$, which is not the unit of a capacitance. On the contrary, they had introduced the term, $C$, as a capacitance in the first paragraph of the Introduction section of their paper as \cite{Fouda2020}:

\textit{``The constant of proportionality between charge and voltage $C=\epsilon A/d$ is the capacitance in
units of Farads.''}

We observe that the authors of Ref.~\cite{Fouda2020} have used the same symbol, $C$, to represent different physical quantities. This makes their treatment and the respective results ambiguous for readers. It is to be noted that the origin of this dimensional inconsistency can be traced back to the paper that is cited as Ref.~[19] in \cite{Fouda2020}, where again the dimension of the Dirac-delta function has been overlooked.

\end{enumerate}

We take this opportunity to clarify that the range of $\alpha$ in Eq.~$\left(13\right)$ of our paper:
\begin{equation}
I_{C_{f}}\left(t\right)=C_{f}\left[\frac{t^{1-\alpha}}{\Gamma\left(2-\alpha\right)}\ast\ddot{V}\left(t\right)\right]=C_{f}\frac{d^{\alpha}}{dt^{\alpha}}V\left(t\right),\label{eq:final_eq}
\end{equation}
is, $0<\alpha<1$, where $C_{f}$ is the pseudocapacitance of the capacitor. Here we prove it through simple mathematical steps. Applying the differentiation property of convolution of two
functions on the two terms that are inside the square bracket of Eq.~$\left(\ref{eq:final_eq}\right)$ above, we have,
\[
\frac{t^{1-\alpha}}{\Gamma\left(2-\alpha\right)}\ast\ddot{V}\left(t\right)=\frac{1}{\Gamma\left(1+1-\alpha\right)}\left[\frac{d}{dt}t^{1-\alpha}\ast\dot{V}\left(t\right)\right].
\]
Differentiating the power-law term inside the square bracket
followed by the use of the functional equation for Gamma functions,
$\Gamma\left(1+z\right)=z\Gamma\left(z\right)$, we have:
\[
\frac{t^{1-\alpha}}{\Gamma\left(2-\alpha\right)}\ast\ddot{V}\left(t\right)=\left[\frac{\left(1-\alpha\right)t^{-\alpha}}{\left(1-\alpha\right)\Gamma\left(1-\alpha\right)}\ast\dot{V}\left(t\right)\right]=\frac{t^{-\alpha}}{\Gamma\left(1-\alpha\right)}\ast\dot{V}\left(t\right),
\]
which when compared with Eq.~$\left(2\right)$ from Ref.~\cite{Pandey2022a}, yields,
\[
\frac{t^{1-\alpha}}{\Gamma\left(2-\alpha\right)}\ast\ddot{V}\left(t\right)=\frac{d^{\alpha}}{dt^{\alpha}}V\left(t\right),\text{ }0<\alpha<1.
\]
Consequently, Eq.~$\left(\ref{eq:final_eq}\right)$, i.e.,  Eq.~$\left(13\right)$ from Ref.~\cite{Pandey2022a}, effectively implies:
\begin{equation}
I_{C_{f}}\left(t\right)=C_{f}\frac{d^{\alpha}}{dt^{\alpha}}V\left(t\right)\text{, }0<\alpha<1.\label{eq:testing}
\end{equation}
Lastly, we convey our thanks to Allagui~et~al. because even though their Comment lacked mathematical consistency, this Response to their Comment has made us even more confident about our results.

%\section*{References}

%\bibliography{mybibfile}

\end{document}